\documentclass{IEEEtran}
\usepackage[T1]{fontenc} 
\usepackage{bm} 
\usepackage{epsfig}
\usepackage{psfrag}
\usepackage{tikz}
\usepackage{array}
\usepackage{graphicx}
\usepackage{multirow}
\usepackage{framed}
\usepackage{float}
\usepackage{caption}
\usepackage{subfig}
\usepackage{tabularx}
\usepackage{url}
\usepackage{color}
\usepackage{balance}
\def\BibTeX{{\rm B\kern-.05em{\sc i\kern-.025em b}\kern-.08em
     T\kern-.1667em\lower.7ex\hbox{E}\kern-.125emX}}

\usepackage{cite}
\usepackage{amsmath,amssymb,amsfonts}
\usepackage{algorithmic}
\usepackage{textcomp}
\usepackage{float}

\begin{document}
\title{Global stability of the Rate Control Protocol (RCP) and some implications for protocol design}
\author{Thomas~Voice, Abuthahir, and~Gaurav~Raina
\thanks{Thomas~Voice is an independent researcher (e-mail: tdvoice@gmail.com).}
\thanks{Abuthahir and Gaurav Raina are with the Department of Electrical Engineering, Indian Institute of Technology Madras, Chennai, India (e-mail: \{ee12d207, gaurav\}@ee.iitm.ac.in).}
}
\maketitle
\begin{abstract}
The Rate Control Protocol (RCP) is a congestion control protocol that relies on explicit feedback from routers.  RCP estimates the flow rate using two forms of feedback: rate mismatch and queue size. However, it remains an open design question whether queue size feedback in RCP is useful, given the presence of rate mismatch. The model we consider has RCP flows operating over a single bottleneck, with heterogeneous time delays. We first derive a sufficient condition for global stability, and then highlight how this condition favors the design choice of having only rate mismatch in the protocol definition.
\end{abstract}
\begin{IEEEkeywords}
Rate control protocol, delay systems, stability of NL systems, communication networks   
\end{IEEEkeywords}
%
\section{Introduction}%
The most widely implemented congestion control algorithm in the Internet today is the Transmission Control Protocol (TCP). Currently, TCP uses packet delay and packet loss as signals of congestion in the network. Delay and loss are both detrimental to packets, and such implicit signaling mechanisms will impose limits on network performance and quality of service. For an excellent review of Internet congestion control, see \cite{srikant2012}. There is interest in congestion control protocols that could utilize more explicit feedback from routers \cite{balakrishnan2007stability,dukkipatircpac,jose2016xcc,katabi2002}. Some well-known examples of explicit congestion control protocols include the eXplicit Control Protocol (XCP) \cite{he2017,katabi2002,liu2012XCPHopf}, Rate Control Protocol (RCP) \cite{balakrishnan2007stability,dukkipatircpac,krbookchap,krv2009,lakshmikantha2008}, MaxNet \cite{Wydrowskimaxnet} and JetMax \cite{Zhangjetmax}.

From a networking perspective, there are currently two architectures that appear appealing as use cases for RCP.  One is a host-centric architecture, which is IP-based, where users and sources communicate with the help of IP addresses. For example, RCP has been explored in a wireless setting \cite{baretto2015rcp}, in data center networks \cite{sharma17}, and also in satellite networks \cite{sun2012rcp}. It is also plausible that users may wish to care only for specific data items; in other words  `what data' is more important than `where data' is. A platform that can enable a user to send a data request without knowing the address of the hosting entity is called Named Data Networking (NDN) \cite{zhangndn2010}. In NDN environments, in order to fetch data, a user sends out an \emph{interest packet}, which carries a name that identifies the data that is desired. NDN uses hierarchically structured names; e.g., a video produced by UCSD may be named /ucsd/videos/WidgetA.mpg. Moreover, data can be fetched from multiple sources, via multiple paths, which makes the implicit signaling mechanism used by TCP unreliable in an NDN setting \cite{ren2016xcc}. For some examples of RCP-style algorithms in NDN environments see \cite{lei15,mahdian2016xcc} and \cite{zhong2017}.

In this paper, our focus will be on one particular design consideration that arises in the protocol definition of RCP. We recall that RCP estimates the flow rate using two forms of feedback: rate mismatch and queue size. However, it remains an outstanding design question whether it is advantageous to include queue size feedback, given that the protocol already includes feedback based on rate mismatch. Currently, regardless of the networking use-cases, the RCP protocol specification uses both rate mismatch and queue size feedback. Some early simulations suggested that incorporating both forms of feedback in RCP may lead to less accurate control over the queue size \cite{krv2009}. However, this insight was based on some initial simulations, and more analysis would be required to arrive at a better understanding of this design choice.

In this paper, we conduct our analysis on a proportionally fair variant of RCP \cite{krbookchap,krv2009}. Rate feedback from RCP routers to end-systems is not instantaneous, so stability in the presence of feedback delays is an important aspect for performance evaluation. In \cite{krv2009}, the authors conducted a local stability analysis, but such an analysis was not able to offer any design insights on the use of two forms of feedback in RCP. In this paper, we resort to a global analysis of RCP and then use the results to infer the potential impact of both design options; i.e., with and without queue size feedback. For some global analysis of different models in the study of Internet congestion control, the reader is referred to \cite{deb2003,hollot2001} and \cite{wang2006}. As far as we know, this is the first global stability analysis of RCP. In our current setting, we consider the RCP model where the flows have heterogeneous time delays, operating over a single bottleneck link. Our analysis offers conditions for global stability which are general enough to incorporate both forms of feedback in RCP. When RCP uses feedback based only on rate mismatch, we get a simply stated condition for global stability that is easier to satisfy, as compared to RCP with two forms of feedback. The results in this paper favor the design choice that uses feedback based only on rate mismatch. However, this is a very important design consideration, and additional analysis and simulations would be needed to arrive at a comprehensive understanding.

The rest of this paper is organized as follows. In Section II, we outline the fluid model of RCP. In Section III, we present results on the global stability of RCP and discuss the implications for protocol design. In Section IV, we outline our contributions and highlight some avenues for further investigation.
\section{RCP Model}
In this section, we recapitulate the proportionally fair variant of RCP introduced in \cite{krv2009}. For other models of RCP, and their analysis, see \cite{balakrishnan2007stability} and \cite{voice2009maxminrcp}. The non-linear fluid model of the system under consideration is \cite{krv2009} 	
\begin{equation}
\label{eq:smallbuf_model}
\dot{R}_{j}(t)=\dfrac{aR_{j}(t)}{C_{j}\overline{T_{j}}(t)}\Big(C_{j}-y_{j}(t)-b_{j}C_{j}p_{j}\big(y_{j}(t)\big) \Big),
\end{equation}
where
\begin{equation}
y_{j}(t) = \sum_{r:j\in r}x_{r}\Big(t-T_{rj}\Big)
\end{equation}
is the aggregate traffic arriving at link $j$. We consider a network with a set $J$ of resources. A route $r$ is a non-empty subset of $J$, and we write $j \in r$ to denote that the route $r$ passes through the link $j$. $R_{j}(t)$ is the fair rate that RCP updates for all flows which traverse through link $j$, $C_{j}$ is the link capacity, $a$ and $b_{j}$ are non-negative protocol parameters. $\overline{T}_{j}(t)$ is given by 
\begin{equation}
\overline{T}_{j}(t) = \dfrac{\sum\limits_{r:j\in r}x_{r}(t)T_{r}}{\sum\limits_{r:j\in r}x_{r}(t)},
\end{equation}
where 
\begin{equation}
 T_{r}=T_{rj} + T_{jr},\quad j\in r.
\end{equation}
Here, $T_{rj}$ and $ T_{jr}$ represent the propagation delay on route $r$ from source to link $j$ and the return delay from link $j$ to source respectively. We assume that the queuing delay can be ignored relative to the propagation delay. The flow rate $x_{r}(t)$ is given by \cite{krv2009}
\begin{equation}
x_{r}(t) = \left( \sum_{j\in r}\Big(R_{j}\big(t-T_{jr}\big)\Big)^{-1}\right)^{-1}.
\end{equation}
A simple approximation for the mean queue size $p_{j}(y_{j})$ is \cite{krv2009}
\begin{equation}
p_{j}(y_{j}) = \dfrac{y_{j}\sigma_{j}^2}{2\Big(C_{j}-y_{j}\Big)},\label{eq:stat_dist}
\end{equation}
where $\sigma_{j}^2$ represents the traffic variability at link $j$. For Poisson traffic, $\sigma_{j}=1$.
Note that the rate equation \eqref{eq:smallbuf_model} contains two forms of feedback: rate mismatch term $C_{j}-y_{j}(t)$, and a term based on the mean queue size.
At equilibrium, for the system (\ref{eq:smallbuf_model}-\ref{eq:stat_dist}) we have
\begin{equation}
 C_j-y_j = b_j C_j p_j\big(y_j\big).
\label{eq:eqbmcond}
\end{equation}
Thus, from (\ref{eq:stat_dist}) and (\ref{eq:eqbmcond}), it follows that at the equilibrium
\begin{equation}
 p^{'}_j\big(y_j\big)=\frac{1}{b_j y_j}.\label{eq:eqbmcond1}
\end{equation}

\section{Global Stability}
In this section, we derive a sufficient condition for global stability of (\ref{eq:smallbuf_model}-\ref{eq:stat_dist}) for a single bottleneck link in the presence of heterogeneous feedback delays. Let $S$ be the set of all routes passing through the bottleneck link. To begin with, we consider a slightly more general system given by 
\begin{equation}
 \dot{R}(t)=\kappa R(t)\Big(f\big(y(t)\big)\Big)^+_{R(t)},\label{eq:rateeqn}
\end{equation}
where $\kappa>0,\ f(\cdot)$ is a strictly decreasing continuously differentiable function, and 
\begin{equation}
 y(t)=\sum_{r\in S} R\left(t-\tau_{r}\right),\label{eq:aggeqn}
\end{equation}
for a set of heterogeneous delays $(\tau_{r})_{r\in S}$. We write $u=(v)^{+}_{w}$ to denote that $u=0$ if $v<0$ and $w\leq 0$, otherwise $u=v$.
We assume that for some $\bar{y}>0$, $ f(\bar{y})=0$. Then the equilibrium rate is given by $\bar{R}=\bar{y}/N$ where $N=\lvert S\rvert$. As before, we let $\bar{T}$ denotes the average of the $\tau_r$ for $r \in S$. We assume $\kappa$ is small enough that $\kappa\bar{T}f(0)<1$. Let $\tau=\max_{r\in S}\tau_r$. We define the following constants,
\begin{equation*}
 w = \frac{\kappa\bar{T}f(0)\bar{y}}{1-\kappa\bar{T}f(0)},\ \quad f^{(1)} = \underset{u \in [0,\bar{y}]}{\max} \frac{f(\bar{y}-u)}{u}.
\end{equation*} 
\begin{equation*}
f^{(2)} = \underset{u \in [0,w]}{\max} -\frac{f(\bar{y}+u)}{u}.
\end{equation*} 

Note, if $f$ is concave then
\begin{equation}
  f^{(1)}=-f^{'}(\bar{y}),\ f^{(2)}=-\frac{f(\bar{y}+w)}{w}.\label{eq:fone} 
\end{equation}
We now have the following result.
\newtheorem{theorem}{Theorem}
\begin{theorem}\label{sufficient theorem}
Suppose that $\kappa$ is chosen such that $\kappa\bar{T}f(0)<1$ and 
\begin{equation}
 \frac{\kappa^2\bar{T}^2\bar{y}(\bar{y}+w)f^{(1)}f^{(2)}}{1-\kappa\bar{T}f(0)}<1, \label{eq:thm}
\end{equation}
then (\ref{eq:rateeqn}-\ref{eq:aggeqn}) is globally asymptotically stable, in that $y(t)\rightarrow\bar{y}$ as $t\rightarrow \infty$, regardless of the initial conditions.
\end{theorem}
\begin{IEEEproof}
Let us define $v(t)=R(t)-\bar{R}$. Let the sets $\check{U}$ and $\hat{U}$ be the set of values such that, for all $\check{u} \in \check{U}$ and $\hat{u} \in \hat{U}$ there exists $T$ such that $v(t) \in [-\check{u}, \hat{u}]$ for all $t>T$.
Since $\bar{R}>0$, it is impossible for $v(t)<-\bar{R}$ for all time. However, once $v(t)\geq-\bar{R}$, from \eqref{eq:rateeqn}, we see that it cannot return below $-\bar{R}$. Thus, we can say $\big[\bar{R},\infty\big)\subset \check{U}$, provided we allow $\infty \in \hat{U}$.

Now, suppose $\check{u} \in \check{U}$ and $T$ is such that $ v(t)\geq-\check{u}$ for all $t>T$. Suppose further that, for some $t>T+2\tau,\ 1 > \epsilon > 0,$ $ v(t)\geq \epsilon v(t^{'})$ for all $t^{'} \in [t-2\tau, t]$. Then for $t^{'}\in [t-\tau,t]$,
\begin{equation*}
 \dot{R}(t^{'})\leq\kappa\Big(\bar{R}+\epsilon^{-1}v(t)\Big)f\left(\bar{y}-N\check{u}\right).
\end{equation*}
Thus, for all $r \in S$,
\begin{equation*}
 R(t)-R\left(t-\tau_r\right)\leq \kappa\Big(\bar{R}+\epsilon^{-1}v(t)\Big)f\left(\bar{y}-N\check{u}\right)\tau_r.
\end{equation*}
Hence,
 $y(t)\geq N \bar{R} + N v(t) - N\kappa\Big(\bar{R}+\epsilon^{-1}v(t)\Big)f\left(\bar{y}-N\check{u}\right)\bar{T}.$
Thus, $\dot{v}(t)<0$ provided
\begin{equation*}
 v(t)> \frac{\kappa\bar{R}f(\bar{y}-N\check{u})\bar{T}}{1-\kappa \epsilon^{-1}f(\bar{y}-N\check{u})\bar{T}}.
\end{equation*}
Thus, $(\hat{u},\infty) \subset \hat{U}$ where
\begin{equation}
\hat{u}= \frac{\kappa\bar{R}f(\bar{y}-N\check{u})\bar{T}}{1-\kappa f(\bar{y}-N\check{u})\bar{T}}.
\label{eq:uhat}
\end{equation}
Setting $\check{u}=\bar{R}$, we get $(w/N,\infty) \subset \hat{U}$. 

Now, suppose $\hat{u} \in \hat{U}$ and $T$ is such that $v(t)\leq\hat{u}$ for all $t>T$. 
Suppose further for some $t>T+2\tau, \ 1>\epsilon > 0$, 
 $v(t)\leq \epsilon v(t^{'}) \, \text{for all} \  t^{'}\in [t-2\tau,t].$ 
Then, for $t^{'}\in[t-\tau,t]$,
\begin{equation*}
 \dot{R}\big(t^{'}\big)\geq \kappa \epsilon^{-1}\big(\bar{R}+\hat{u}\big)f(\bar{y}+N\hat{u}).
\end{equation*}
Thus, for all $r \in S$,
\begin{equation*}
 R(t)-R(t-\tau_r)\geq\kappa \epsilon^{-1}\big(\bar{R}+\hat{u}\big)f(\bar{y}+N\hat{u})\tau_r.
\end{equation*}
Hence, 
 $y(t)\leq N \bar{R} + N v(t) - N \kappa \epsilon^{-1} \big(\bar{R}+\hat{u}\big)f(\bar{y}+N\hat{u})\bar{T}.$
Therefore, $\dot{v}(t)>0$ provided
\begin{equation*}
 v(t) < -\kappa \epsilon^{-1}\big(\bar{R}+\hat{u}\big)\vert f(\bar{y}+N\hat{u})\vert \bar{T}. 
\end{equation*}
Thus, $(\check{u}, \infty) \subset \check{U}$ where,
\begin{equation}
\check{u}=\kappa\big(\bar{R}+\hat{u}\big)\vert f(\bar{y}+N\hat{u})\vert \bar{T}.
\label{eq:ucheck}
\end{equation}
Now, the right hand side of \eqref{eq:ucheck} is continuous in $\hat{u}$, and 
\begin{align*}
 \kappa \frac{(\bar{y}+w)}{N}\bar{T}\vert f(\bar{y}+w)\vert &\leq \kappa \bar{T}(\bar{y}+w)\frac{w}{N} f^{(2)},\\ \nonumber
 &=\frac{\kappa^2 \bar{T}^2 (\bar{y}+w)f^{(2)}f(0)\bar{R}}{1-\kappa\bar{T}f(0)} ,\\
 &< \bar{R},  \nonumber
\end{align*}
from \eqref{eq:thm}. Thus 
$\big(\bar{R}-\delta, \infty\big) \subset \check{U} \, \text{for some}\, \delta >0$. Hence, $\Big((w/N)-\delta^{'}, \infty\Big) \subset \hat{U}$ for some $\delta^{'}>0.$
Now, for any $\check{u} \in \check{U}$, from \eqref{eq:uhat}, $(\hat{u}, \infty) \subset \hat{U}$ for 
\begin{equation*}
 \hat{u}=\frac{\kappa\bar{T}\bar{y}f^{(1)}\check{u}}{1-\kappa f(0)\bar{T}}.
\end{equation*}
Let $\hat{v}=\min\big(\hat{u}, (w/N)-\delta^{'}\big)$. From \eqref{eq:ucheck}, we know that
 $(\check{v},\infty) \subset \check{U}$,
where 
\begin{equation*}
 \check{v}=\kappa\bar{T}(\bar{y}+w)f^{(2)}\hat{v}\leq \frac{\kappa^2\bar{T}^2\bar{y}(\bar{y}+w)f^{(1)}f^{(2)}\check{u}}{1-\kappa\bar{T}f(0)}.
\end{equation*}
Thus, from \eqref{eq:uhat}, $\check{v}\leq\gamma\check{u}$ for $\gamma < 1$. Therefore, $\check{U}=(0,\infty)$, and so, from \eqref{eq:uhat}, $\hat{U}=(0,\infty)$. This implies that $v(t)\rightarrow0$ as $t \rightarrow \infty$. As our initial conditions were arbitrary, (\ref{eq:rateeqn}-\ref{eq:aggeqn}) must be globally asymptotically stable, as required.
\end{IEEEproof}
To highlight the implications of {Theorem \ref{sufficient theorem}}, we now apply these results to analyze the global stability of RCP by choosing appropriate functions for $f(\cdot)$.\\ \\
\textbf{Case A} (RCP with queue feedback): Suppose we use the function given in \eqref{eq:smallbuf_model},
\begin{equation*}
 f(y)=C-y-\frac{bC\sigma^2y}{2(C-y)},
\end{equation*}
for constants $C$, $b$ and $\sigma$.

As in \eqref{eq:smallbuf_model}, we let $a = \kappa C \bar{T}$. In this case, $f(\cdot)$ is concave, so \eqref{eq:fone} applies. Now, for this $f(\cdot)$,  $w=a\bar{y}/(1-a).$ So, with appropriate choice of $a$ we can make $w$ arbitrarily small. In order for $f^{(2)}$ to be bounded, we need $\bar{y}+w<C$. Indeed, since $f(y)\rightarrow\infty$ as $y\rightarrow C$, we need $y(t)<C$ for all $t$ in order for solutions to be well defined. However, from the above proof, this will be true if the condition  \eqref{eq:thm} is satisfied and $y(t)<\bar{y}+w<C$ for $t\leq0$. Now, suppose we choose $\kappa$ so that
 $w=\alpha(C-y) \, \text{for some} \,  \alpha<1.$
Then, from \eqref{eq:fone},
\begin{equation*}
 f^{(2)}=\frac{\vert f(\bar{y}+w)\vert}{w} = \frac{1-\alpha}{\alpha} + \frac{bC(\bar{y}+w)\sigma^2}{2(1-\alpha)\alpha(C-\bar{y})^2}.
\end{equation*}
At equilibrium, $2(C-\bar{y})^2 = bC\bar{y}\sigma^2$. Thus,
\begin{equation*}
 f^{(2)} = \frac{1-\alpha}{\alpha} + \frac{1}{(1-\alpha)\alpha} + \frac{bC\sigma^2}{2(1-\alpha)(C-\bar{y})}.
\end{equation*}
But 
 $C=\bar{y}+(C-\bar{y})\leq \sqrt{C\bar{y}} + (C-\bar{y}),$
hence,
\begin{equation*}
 f^{(2)}\leq \frac{1-\alpha}{\alpha} + \frac{1}{(1-\alpha)\alpha} + \frac{\sqrt{2b}\sigma + b\sigma^2}{2(1-\alpha)}.
\end{equation*}
Furthermore, from \eqref{eq:eqbmcond1} and \eqref{eq:fone}, we get $f^{(1)}=1+C/y$.
Now $\bar{y}+w = \bar{y}/(1-a)$. Thus, from \eqref{eq:thm}, stability is guaranteed if
\begin{equation*}
 1>\frac{\kappa^2\bar{T}^2\bar{y}^2}{(1-a)^2}f^{(1)}f^{(2)}=\frac{w^2}{C^2}f^{(1)}f^{(2)}.
\end{equation*}
Hence, a sufficient condition for stability is,
\begin{equation*}
 1> \alpha^2 \frac{(C-\bar{y})^2}{C^2} \times \frac{2C}{\bar{y}} f^{(2)}=\alpha^2b\sigma^2f^{(2)}.
\end{equation*}
This will hold if $\alpha<\alpha^{'}$ where
\begin{equation}
 \alpha^{'}=\min\Bigg(\frac{1}{2}, \frac{2}{b\sigma^2\big(6+\sqrt{2b}\sigma+b\sigma^2\big)}\Bigg).\label{eq:alphadash}
\end{equation}
In order for $w<\alpha^{'}(C-\bar{y})$, we need,
 ${\kappa\bar{T}C\bar{y}}/(1-\kappa\bar{T}C) < \alpha^{'}(C-\bar{y})$,
which is implied by,
\begin{equation*}
 \frac{a}{1-a} < \alpha^{'}\sigma\sqrt{\frac{bC}{2\bar{y}}}.
\end{equation*}
Now, 
 $C/\bar{y} = b\sigma^2 C^2/2(C-\bar{y})^2 \leq \max(1,b\sigma^2/2).$
Hence we have global asymptotic stability, provided,
\begin{equation}
 \alpha < \frac{ \alpha^{'}\max\left(\sqrt{2b}\sigma,b\sigma^2\right) } {2 + \alpha^{'}\max\left(\sqrt{2b}\sigma,b\sigma^2\right)},\label{eq:gscondwithq}
\end{equation}
where $\alpha^{'}$ is given in \eqref{eq:alphadash}.\\ \\ 
\textbf{Case B} (RCP without queue feedback): Now, let us consider the special case where $b=0$ in \eqref{eq:smallbuf_model}.

Again, we let $a=\kappa C\bar{T}$. So, $f(y)=C-y$, $f^{(1)}=f^{(2)}=1$, $\bar{y}=C$ and $\bar{y}+w = \bar{y}/(1-a)$.
Hence {Theorem \ref{sufficient theorem}} applies, provided $a<1$ and,
\begin{equation}
 \frac{a^2}{(1-a)^2} < 1.
\end{equation}
Thus $a<1/2$ is a \textit{sufficient} condition for global asymptotic stability. Note that the stability condition obtained in this case is easier to satisfy, as compared to that of RCP with two forms of feedback.\\

\textbf{Discussion:}
If we compare the conditions for global stability for the cases $b=0$ and $b>0$, we find an interesting difference. If we let $b\rightarrow 0$, then $a$ must scale with $\sqrt{b}$ in order to satisfy the stability condition, however at $b=0$, $a<1/2$ is sufficient. The reason for this is that if $b$ is small, then $\bar{y}$ is close to $C$ and $f^{'}(y)$ increases rapidly as $y$ varies from $\bar{y}$ to $w$. This means that $a$ must be small, otherwise the algorithm progresses too quickly. However, if $b=0$, then it is no longer necessary for $y(t)<C$ for all $t$, and $f^{'}(y)$ is constant, and the condition on $a$ is much less strict. If $b$ becomes large, then $a$ will have to scale with $b^{-1}$ in order to satisfy \eqref{eq:gscondwithq}, but $\vert f^{'}(y)\vert$ will roughly scale with $b$, for all $y<C$. So in terms of the derivative of $\kappa f(y)$, these two effects should cancel each other out. However, as $b$ becomes larger, $\bar{y}$ and $\bar{R}$ become smaller, and this will have a slowing effect on 
the rate of convergence of the algorithm, since \eqref{eq:rateeqn} has an $R(t)$ term. Note, if $b$ is very small, and $\bar{y}$ is very close to $C$, then limitations on buffer length will means that $\sigma^2y/2(C-y)$ is no longer a good model for the average queue size. Indeed, if $b$ is small compared to the inverse of the maximum buffer size at the link, then this model breaks down, and it is more appropriate to simply remove the queue term from $f(\cdot)$.

Another interesting comparison can be made between the conditions for \textit{local} and \textit{global} stability of RCP. For the RCP dynamical system which uses both rate mismatch and queue size feedback, a sufficient condition for local asymptotic stability is \cite{krv2009}
\begin{equation}
a < \frac{\pi}{4}.
\label{krveq:cond}
\end{equation}
In contrast to the condition for global stability, as outlined in \eqref{eq:gscondwithq}, this simple sufficient condition for local stability places no restriction on the value of the parameter $b$. If the parameter $b$ is set to zero, one may also derive an alternative less-conservative sufficient condition for local stability. In this case, a sufficient condition for the RCP system to be locally stable about its equilibrium is \cite{krv2009}
\begin{equation}
a < \frac{\pi} {2}.
\label{krveq:cond2}
\end{equation}
Our sufficient condition for global stability is $a<1/2$ which is more restrictive than the local stability condition \eqref{krveq:cond2}. However, this condition for global stability is rather attractive as it does not depend on any other network parameters like the feedback delay of the RCP flows, or the link capacity.                

\balance
\section{Contributions}
The motivation for our study was to get a better understanding of a key design choice in RCP; i.e., whether queue size feedback in RCP is useful, given the presence of rate mismatch. In this paper, we first constructed a sufficient condition to ensure global asymptotic stability for a non-linear delay differential equation with heterogeneous delays. Then, we employed the results to RCP to find that the removal of queue feedback yields a condition that is easier to satisfy. The analytical results favor the design choice that uses only rate mismatch feedback. Our work is a contribution to the global analysis of RCP and also to a key design consideration that is currently present in the RCP protocol definition.  

Other ways to develop additional insight into RCP would be to consider a local bifurcation analysis, and also to consider a dynamic network environment where users arrive and depart in an RCP network.

\end{document}